\documentclass[hyperref,solaromanenum,lines,longbibliography,linenumbers]{spr-sola}

\usepackage{ifluatex}
\ifluatex
\usepackage{pdftexcmds}
\usepackage{color}  
\usepackage{lineno}                   
\makeatletter
\let\pdfstrcmp\pdf@strcmp
\let\pdffilemoddate\pdf@filemoddate
\fi
\usepackage{comment}
\usepackage{soul}
\usepackage{ulem}
\usepackage[figuresright]{rotating}
\usepackage{hyperref}
\usepackage{newunicodechar}
\newunicodechar{₃}{$_3$}

\usepackage{soul}

\usepackage{graphicx}                    

\usepackage{amsmath}
\usepackage{booktabs}
\usepackage{float}
\chardef\us=`\_
\graphicspath{{./}{figures/}}

\begin{document}

\begin{frontmatter}

\title{Spectroscopic \ion{He}{1} 1083 nm prominence eruption observations in the middle corona with MLSO/UCoMP}
\author[addressref={aff0,aff1}]{\inits{C.}\fnm{Chloe}~\snm{Pistelli}}
\author[addressref={aff1},email={momchil.molnar@swri.org}]{\inits{Corresponding author: M.}\fnm{Momchil E.}~\snm{Molnar}\orcid{0000-0003-0583-0516}}
\author[addressref={aff2}]{\inits{G.}\fnm{Giuliana}~\snm{de Toma}\orcid{0000-0002-8439-6166}}
\author[addressref={aff1}]{\inits{J.}\fnm{Joseph}~\snm{Plowman}\orcid{0000-0001-7016-7226}}

\address[id=aff0]{Taylor University, Upland, IN, 46989, USA}
\address[id=aff1]{Southwest Research Institute, Boulder, CO, 80301, USA}
\address[id=aff2]{High Altitude Observatory, 
	NSF National Center for Atmospheric Research, Boulder, CO, 80301, USA}



\begin{abstract}

Coronal mass ejections (CMEs) are a major driver of space weather as they propagate through the heliosphere. Many CMEs have associated prominence material entangled in their magnetic structure which contains cooler plasma. This cooler CME component contains significant amounts of neutral elements, which emit brightly in permitted atomic lines. It has been hypothesized that permitted transitions of neutral elements in eruptions could be used for inferring the magnetic field in CMEs, which is crucial for space weather forecasting. We present observations made with the Upgraded Coronal Multi-channel Polarimeter (UCoMP) in \ion{He}{1} 1083\,nm that clearly show the presence of neutral helium in eruptive prominences associated with CMEs  as they propagate through the lower and  middle corona. 
We find that solar prominence eruptions can be detected in \ion{He}{1} 1083\,nm observations up to the edge of the instrument field of view at $\approx$\,2 solar radii, providing valuable spectral information that complements existing extreme ultraviolet and white-light coronal imaging observatories. These results illustrate  the capability of UCoMP to probe the dynamic behavior of prominence eruptions, allowing for their line-of-sight velocity estimation, and potentially improving space weather forecasting by enabling earlier and more accurate identification of Earth-directed eruptions.
\end{abstract}

    \end{frontmatter}


\section{Introduction} 
Solar prominences are large-scale magnetic structures composed of cool ($\approx10^4$\,K), dense plasma seen over the solar limb (or filaments if seen on the disk). They extend into the corona, where they are supported
by shared or twisted magnetic field configurations \citep[see][review and references therein]{2018LRSP...15....7G}. Prominences exhibit a multitude of appearances and can be found at different latitudes on the solar disk, with differing lifetimes, while always forming above polarity inversion lines  \citep{Parenti2014Prominences}. When their magnetic equilibrium configuration is disrupted they can erupt and be entrailed in a coronal mass ejection (CME). Such eruptive events represent a primary driver of space weather and can generate geomagnetic storms with significant implications for astronauts, spacecraft, and technological infrastructure on Earth~\citep{2021LRSP...18....4T}.
In prominences the unsaturated Hanle effect could be successfully used to infer the (line-of-sight) magnetic field of erupting flux ropes~\citep{Molnar_2024}, which is the missing link in coronal magnetic field observations currently~\citep{2017SSRv..210..145C}. Consequently, observing the \ion{He}{1} 1083\,nm line is a promising avenue for readily available ground-based measurements of the magnetic field in the CME prominence core.

However, one of the unresolved challenges of CME magnetic measurement is how far above the solar limb is neutral helium detected in the entrained erupting prominence material. Previous observations with the Solar Maximum Mission~\citep{1980SoPh...65....5B} have shown prominence eruptions extending to more than 3\,R$_{\odot}$ above the solar limb in \ion{H}{1} 656.3\,nm (H-$\alpha$) line~\citep{1979SoPh...61..201M,1981ApJ...244L.117H,1986SoPh..105..173I}. The Mauna Loa Solar Observatory Chromospheric Helium-I Imaging Photometer (MLSO/CHIP) instrument detected prominence eruptions up to the edge of its FOV at 1.4\,R$_{\odot}$~\citep{2004SoPh..225..337M}. To answer this question, we use data from the commissioning phase of the Upgraded Coronal Multichannel Polarimeter \citep[UCoMP][]{2019shin.confE.131T} to study the evolution of prominences during prominence eruptions. UCoMP is a ground-based coronagraphic spectropolarimeter located at the Mauna Loa Solar Observatory (MLSO) on Hawaii and is operated by the NSF National Center for Atmospheric Research (NCAR) High Altitude Observatory (HAO). UCoMP is specifically designed to perform high-precision polarimetric measurements of the solar corona with full Stokes polarimetry between 530\,nm to 1083\,nm~\citep{2008SoPh..247..411T}. The instrument is capable of spectropolarimetric scans across a broad range of coronal emission lines within the visible to near-infrared wavelengths (520-1083\,nm), enabling the study of the coronal magnetic fields~\citep{2020Sci...369..694Y}, the ubiquitous coronal Alfv\'en waves \citep{2007Sci...317.1192T, 2016ApJ...828...89M}, and the coronal thermal plasma properties during eruptions~\citep{2013SoPh..288..637T,
2023SPD....5420501W, 2023FrASS..1092881S}.

The low sky brightness at the Mauna Loa Observatory and the unique coronographic design of UCoMP allows it to observe regularly the solar corona, which is typically a few millionths of the disc center brightness. We use this unique capability of UCoMP to study the propagation of prominence eruptions seen in \ion{He}{1} 1083 through the low and middle corona, for the first time. The aim of this paper is to establish the observability of the prominences with the UCoMP instrument (and coronographs in general) and make a foundational estimate of their suitability for space weather forecasting diagnostics of ICMEs. We describe the UCoMP \ion{He}{1} 1083 observations in Section~\ref{sec:Observations}; we describe the prominence eruption properties found in the data archive in Section~\ref{sec:Results}, and then in Section~\ref{sec:discussion} we discuss the prospect forward based on our results.

\section{Observations from the UCoMP Commissioning Phase}
\label{sec:Observations}

We describe below the observations taken during the commissioning phase of the new UCoMP instrument in the \ion{He}{1} 1083 line taken between July 2021 and January 2022. This data set contains 79 observing days with \ion{He}{1} 1083 data. After examining the observations we found prominences extending  significantly above the UCoMP occulter (at $\approx$1.04-1.05\,R$_{\odot}$)
on 14 out of the 79 observing days. Only a subset of these prominence observations are presented here, because most of these events were observed in a single image or under conditions of unfavorable seeing or instrumental settings. UCoMP stopped observing \ion{He}{1} 1083 in February 2022, so no further data were available to us.

\begin{figure*}[htbp]
    \centering
    \includegraphics[width=1.\linewidth]{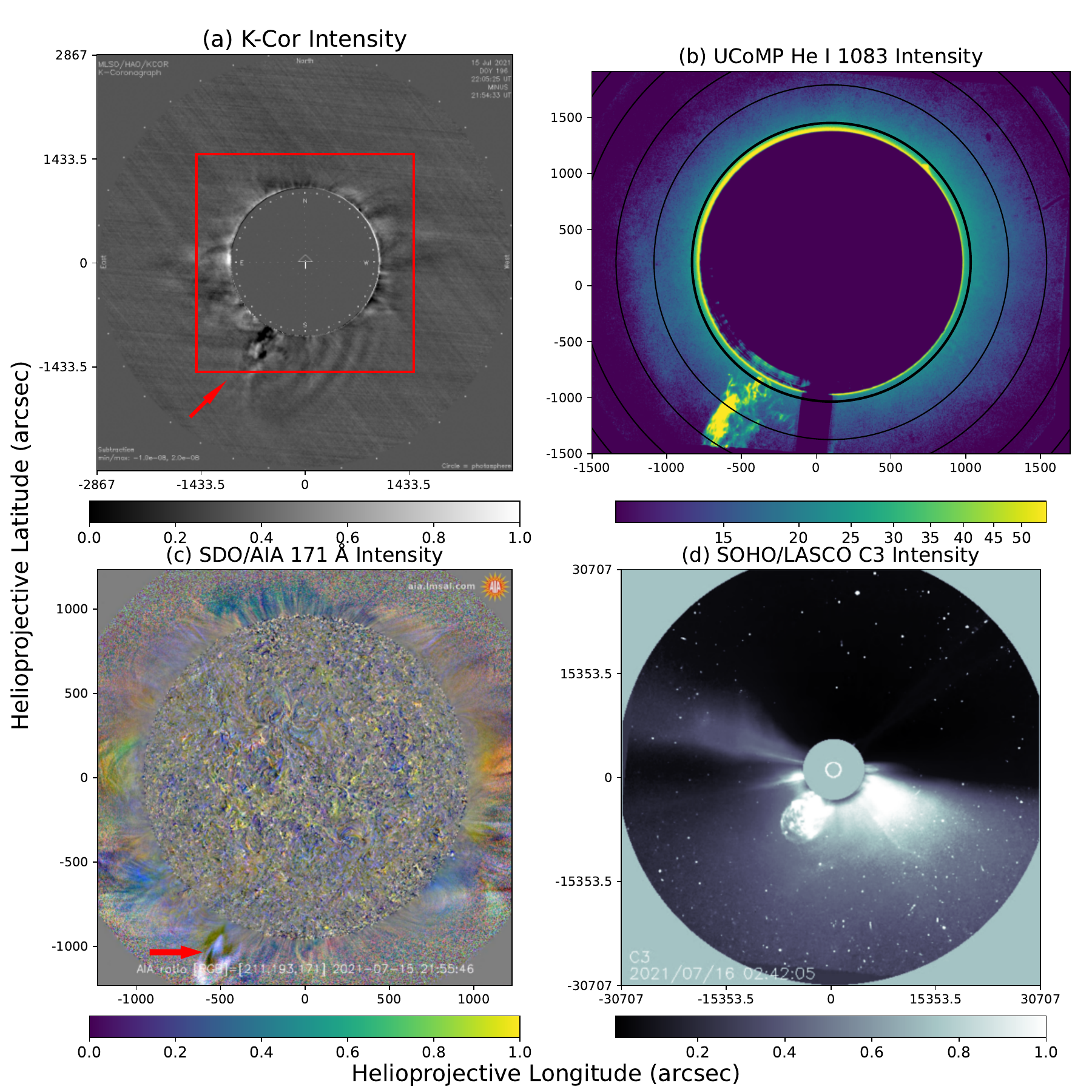}

    \caption{A context illustration of the solar eruption on 2021 July 15; \emph{panel (a)}: MLSO/K-COR white light coronagraph difference image at 22:05:25 UT; \emph{panel (b):} UCoMP \ion{He}{1}1083 line core intensity at 21:53:22 UT;
    \emph{panel (c):} SDO/AIA 211,193,171\,{\AA} RGB composite from the AIA LMSAL webpage at 21:55:46 UT; 
    \emph{panel (d):} SOHO/LASCO C3 coronograph white light image at 02:42:05 UT on 2021 July 16. The red arrow in the K-COR and SDO/AIA plots point to the observed eruption. The red rectangular box in the K-Cor plot represents the field of view of UCoMP.}
    \label{fig:K-Cor_AIA_UCoMP_Plot}
\end{figure*}

\subsection{UCoMP Data}
\label{subsec:UCoMP_issues}

The UCoMP instrument is an imaging spectropolarimeter. It consists of two cameras to simultaneously record an emission line and the continuum nearby. The continuum channel allows to remove the sky and continuum contribution from the actual emission line. This subtraction is critical for a proper calibration of the data. 
The instrument transmission profile is shown in Figure 3. The wavelength offset of the off-band continuum channel is set by the properties of the Lyot filter and cannot be adjusted. In the case of the bright \ion{He}{1} 1083 line a secondary lobe from the on-band channel contaminate the off-band continuum channel. Therefore, a proper background correction cannot be performed for this line without first removing this spurious contribution, and such correction, while possible, is not trivial and require a realistic instrument model. In this analysis, we were only interested in intensity data, and because the prominence material is significantly brighter than the sky background, this correction was expected to be small. Our approach was to estimate an average intensity background as a function of height using the region around the prominence location (please note that this approach will not work for polarization data). 
We found a background intensity level between 10-15\,$\mu$B$_{\odot}$ decreasing away from the occulter, consistent across wavelengths. We found a small variation on the order of 10$\%$ between different observation times and dates. The measured sky background level is significantly lower than the erupting prominence signal in the data and will not be further treated in this publication, but will be important for future detailed studies of the UCoMP \ion{He}{1} data. 

The \ion{He}{1} 1083 data taken by UCoMP during its commissioning phase in 2021-2022 are not publicly available because the instrument was not optimized to observe this bright emission line and \ion{He}{1} observations suffered from several issues, as discussed below. We obtained the \ion{He}{1} observations via a data request to the MLSO team, browsed through the data, and marked all possible candidate observations for prominence eruptions.
We cross checked the prominence eruptions with the ones detected in the SoHO LASCO~\citep{1995SoPh..162..357B} CME catalog~\citep{2004JGRA..109.7105Y} as further discussed in Section~\ref{sec:discussion}. An example of a well observed prominence eruption in the \ion{He}{1} 1083 data from the 2021 July 15 CME is shown in Figure~\ref{fig:K-Cor_AIA_UCoMP_Plot}, where data from other observatories provide context imaging for this eruption. The occulter of UCoMP is shown as the dark circle in the center, that has a size of about 1.04\,R$_{\odot}$. This halo CME is associated with a flare beyond the east limb which occurred around 2021 July 15 21:15 UT, with first appearance in the LASCO C2 field of view (FOV) at 21:36:05 UT. We also found well captured CMEs on the dates of
2021 July 15 and 30 and 2021 August 28. The observations taken during these periods had spectra measured at 7 wavelength positions -- 1082.64, 1082.76, 1082.88, 1083.0, 1083.12, 1083.24, and 1083.36\,nm.

Another challenge to derive intensity in prominences, was the much higher intensities of prominences in the \ion{He}{1} 1083 line (up to 1000\,$\mu$B$_{\odot}$) compared to regularly observed coronal lines, such as \ion{Fe}{13} 1074 and 1079\,nm that have intensities of the order of 10\,$\mu$B$_{\odot}$ or less.
To account for these larger intensities, the UCoMP cameras were operated in the \ion{He}{1} 1083 line with a shorter exposure time and/or a different gain during the commissioning phase. Despite adjusting the setting of the cameras, many \ion{He}{1} 1083 observations were in the non-linear or saturation regime of the cameras. Hence, any intensity values provided in this paper should be considered a lower estimate of the real prominence intensity.

\begin{figure*}
    \centering
    \includegraphics[width=0.8\linewidth]{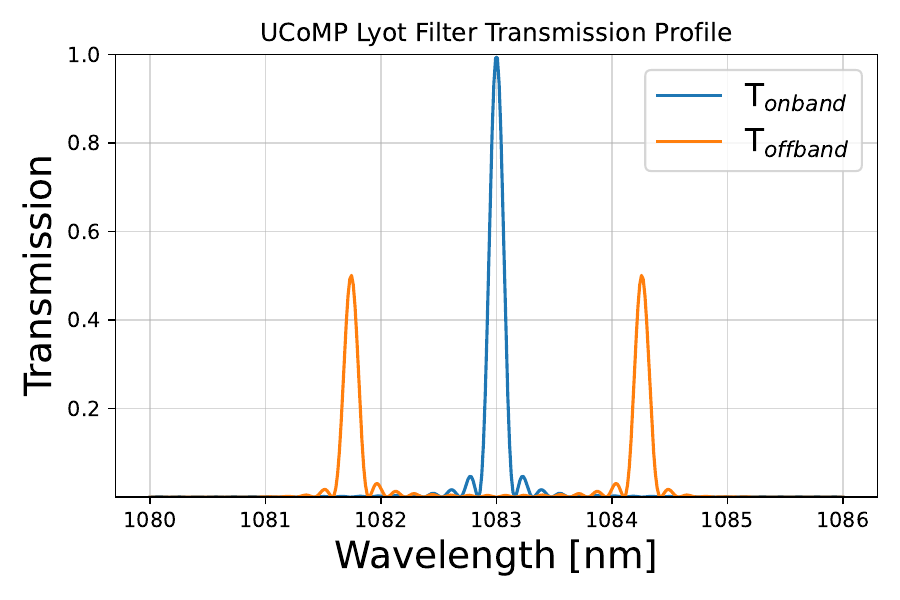}
    \caption{Transmission of the UCoMP Lyot filter in the \ion{He}{1} 1083 window. The on-band channel corresponds to the blue curve with a single peak, whereas the off-band channel, used for the background subtraction, is shown in orange and has a double peaked profile.}
    \label{fig:UCoMP_transmission}
\end{figure*}

In summary, the UCoMP data from the commissioning phase of the instrument suffers from several issues and it was not possible to fully calibrate the intensity images, nor to use the polarimetric observations. Nevertheless, as shown in section \ref{sec:Results}, these issues are not insurmountable for proving the existence and utility of prominence emission in the middle corona. The rather bright prominence material is clearly seen in the intensity images and can be qualitatively studied from these data.

\subsection{Context imaging during the UCoMP commissioning campaign: K-Cor and AIA}

To provide context images of the solar corona during the commissioning phase of UCoMP, we examined the data from two instruments -- K-Cor and AIA, described below. They provide multithermal context imaging of the corona, which complements the UCoMP \ion{He}{1} 1083 data.

The K-coronagraph (K-Cor) is an internally occulted coronagraph at the MLSO that is designed to study the formation and dynamics of coronal mass ejections and the evolution of the density structure of the corona~\citep{deWijn_2012}. The K-Cor records the polarization brightness (pB) formed by Thomson scattering of photospheric light by coronal free electrons. It observes in a synoptic fashion and provides regular coronal observations during periods of good atmospheric conditions. In our case, as described below, K-Cor provided imaging of the front and core of CMEs extending the FOV obtainable with UCoMP up to 3\,$R_{\odot}$. Figure~\ref{fig:K-Cor_AIA_UCoMP_Plot} shows a comparison between the UCoMP and K-Cor data products, emphasizing the complementary information from UCoMP to whitelight CME observations.

\begin{sidewaysfigure}[htp]
    \centering
    \includegraphics[width=\linewidth]{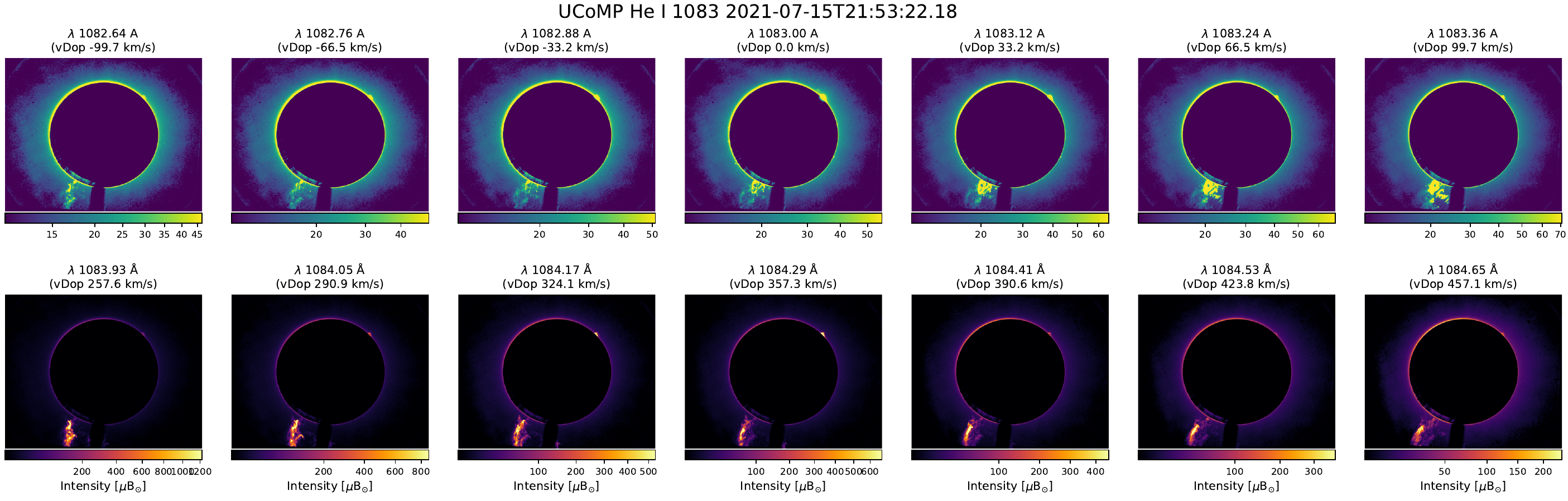}
    \caption{Intensity of the prominence eruption on 2021 July 15 across the \ion{He}{1} 1083 spectral line in units of millionths of disc brightness ($\mu$B). The top row shows the on-band data, and the bottom row shows the off-band continuum channel. The Doppler velocities in the bottom panel of Figure 3 correspond a wavelength shift equal to the difference between the \ion{He}{1}1083 line peak and the wavelength at which the off-band image was tuned. They are intended to be a guide to the reader and we do not imply them to be present in the emitting plasma. Due to the non-linear camera detection regime of the bright prominence, the intensity estimates are lower bounds of the real intensities.}
    \label{fig:amp_figure_20210715}
\end{sidewaysfigure}

\begin{figure}[htp]
    \centering
    \includegraphics[width=\linewidth]{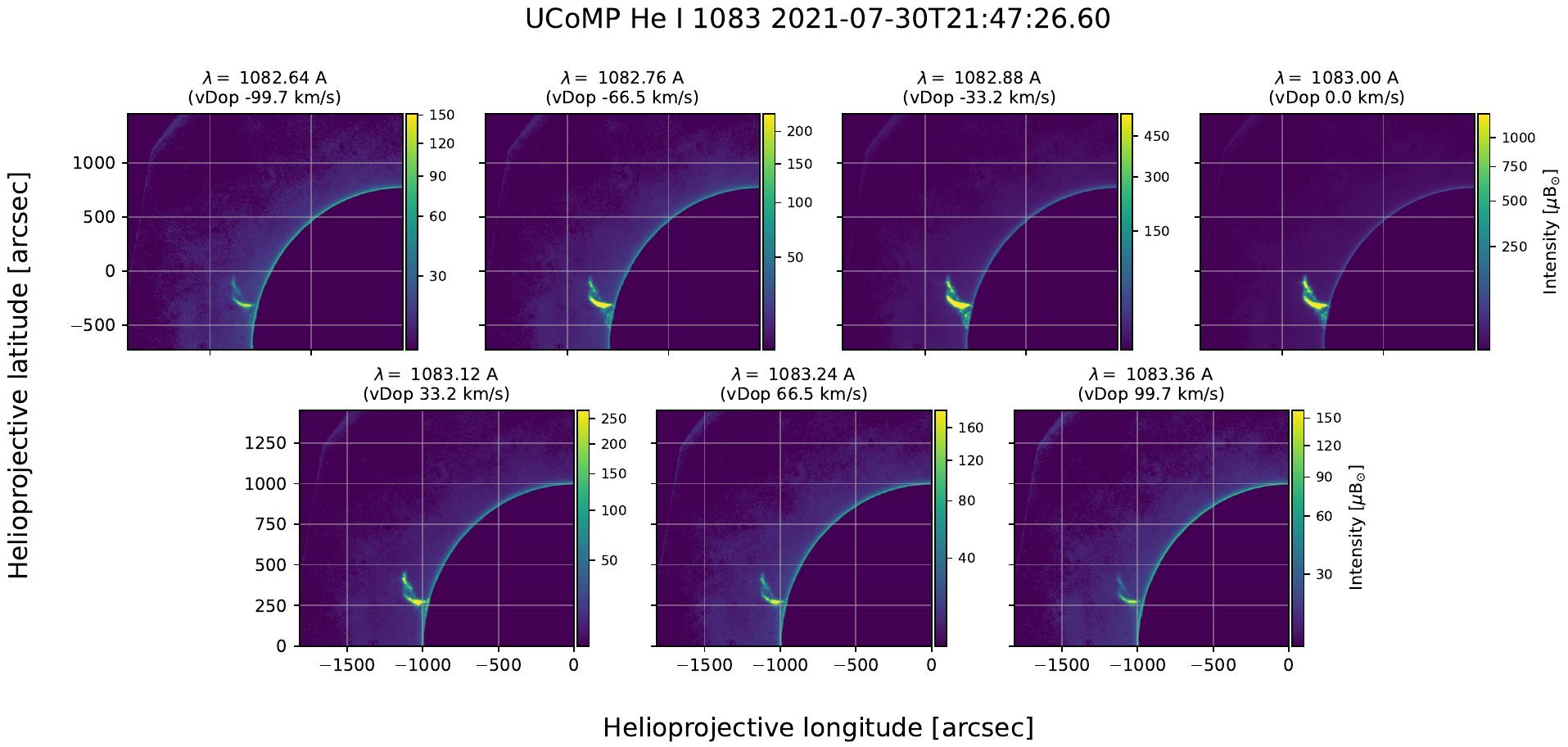}
    \caption{Intensity of the solar eruption on 2021 July 30 across the \ion{He}{1} 1083 spectral line window in units of millionths of disc brightness ($\mu$B). The same caveats for the data presented in Figure~\ref{fig:amp_figure_20210715} apply.}
    \label{fig:amp_figure_20210730}
\end{figure}

\begin{figure}
    \centering
    \includegraphics[width=\linewidth]{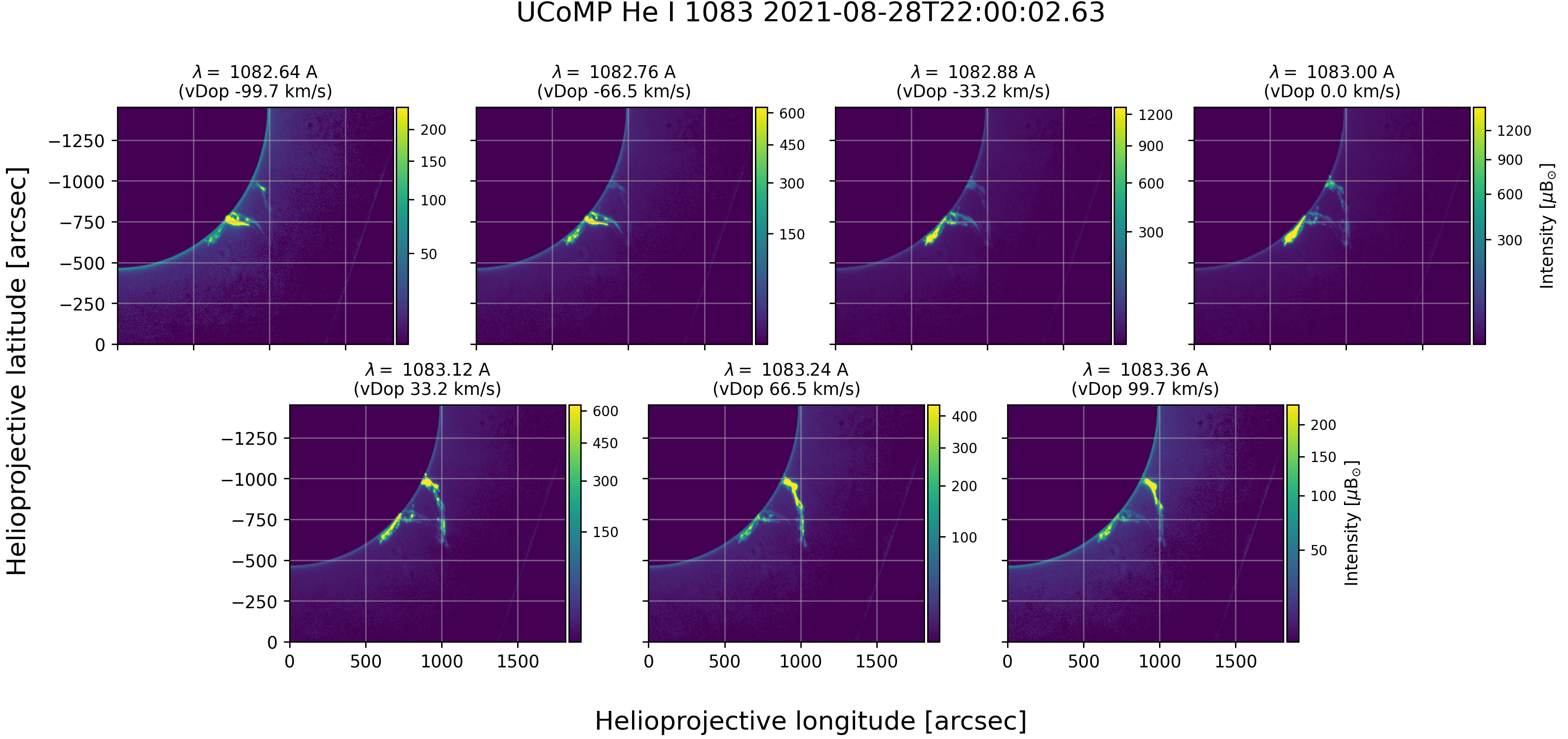}
    \caption{Same as Figure \ref{fig:amp_figure_20210730} for the prominence eruption observed on 2021 August 28.}
    \label{fig:amp_figure_20210828}
\end{figure}

We combined the UCoMP \ion{He}{1} 1083\,nm data with observations from the Atmospheric Imaging Assembly \citep[AIA,][]{AIA_paper} onboard NASA’s Solar Dynamics Observatory \citep[SDO,][]{2012SoPh..275....3P} to obtain a multithermal view of the solar plasma conditions in the observed CMEs. Our objective was to determine the connection between the cool plasma seen in the \ion{He}{1} 1083 data and the hotter coronal plasma observed with SDO/AIA. Once eruptions were located in the UCoMP data, we focused specifically on the 304\,{\AA} and 171\,{\AA} passbands, which are well-suited for capturing emission from plasma with cooler temperatures (between 50,000\,K to 10$^6$\,K).
To improve the detectability of the faint CME structures in the AIA data, we employed several image enhancement techniques. In the AIA data, we applied the \texttt{sunpy}~\citep{sunpy_community2020} implementation of the Normalizing Radial Gradient Filter \citep[NRGF,][]{2006SoPh..236..263M} which suppress the steep radial intensity drop-off of the solar corona, thereby enhancing faint off-limb features.

\section{Properties of prominence eruptions as observed in the \ion{He}{1}1083 line}
\label{sec:Results}

The full set of images at all wavelengths observed with UCoMP in the \ion{He}{1} 1083 channel for the eruptions seen on 2021 July 15 and 31 and August 28 are shown in Figures~\ref{fig:amp_figure_20210715}, \ref{fig:amp_figure_20210730}, and \ref{fig:amp_figure_20210828}, respectively. 
In these figures, we show the on-band emission line observations of UCoMP, 
 (see Figure~\ref{fig:UCoMP_transmission}, blue transmission profile).
For the fast event on 2021 July 15, we also show the corresponding off-band continuum images in the bottom rows
(see Figure~\ref{fig:UCoMP_transmission}, orange transmission profile).
As a guide to the reader, for each image, we have noted the Doppler velocity that would correspond to a wavelength shift equal to the difference between the \ion{He}{1} 1083 line center and the wavelength at which the filter was tuned, where we do not 
imply such particular velocities to be present in the emitting plasma.

The observations of the prominence eruption of 2021 July 15 in Figure~\ref{fig:amp_figure_20210715} show the eruption extending beyond the edge of the field of view with signal present across all wavelength channels. Given the caveats of the UCoMP data described in Section~\ref{subsec:UCoMP_issues}, we see that the emission becomes brighter in the on-band images with increasing wavelength. This indicates a clear redshift of the erupting prominence. We, thus, can assume that the signal in the off-band images originates from the redshifted transmission peak of the off-band channel (see orange transmission profile in Figure~\ref{fig:UCoMP_transmission}). We note that the off-band images show intensities as high or higher than in on-band data. Additionally, parts of the eruptive prominence are only visible, and very bright, in the off-band images. This strong prominence signal in the off-band images is a clear indication of velocities on the order of a few hundred km/s away from the observer in this part of the CME prominence core.

The observations from UCoMP provided us with sparse time estimates of the line-of-sight velocity of the erupting prominence. We compared it with the plane of the sky velocity derived from the K-Cor data to show the complementary nature of these observations, as presented in Figure~\ref{fig:K-Cor_vel_plot}. The K-Cor velocity estimates were derived by locating the position of the prominence / CME core in the K-Cor NRGF data product; the presented uncertainties were estimated from the uncertainty of the estimation of the the CME core seen in the K-Cor data. We find an average speed of propagation of CME core of about $\sim$420\,km/s, which is about half of the one observed in the CME front in this eruption (Burkepile, J., private communication). The propagation speed of the CME core is consistent with the Doppler shifts seen in the off-band UCoMP data, previously seen in Figure~\ref{fig:amp_figure_20210715}.  

We, thus, assume that the signal originates from the redshifted transmission peak of the off-band channel (orange transmission peak in Figure~\ref{fig:UCoMP_transmission}) and indicate associated redshifted velocities of a few hundred kilometers away from the observer of this part of the eruptive prominence. Comparing this with the inferred CME front velocity from the K-Cor data, shown in Figure~\ref{fig:K-Cor_vel_plot}, we see an agreement between the two measurements. However, we find that the line-of-sight and plane-of-sky velocities of this event indicate a fast CME. The large line-of-sight velocity component detected by UCoMP is in agreement with the LASCO observations which report this as a fast halo event reaching over 1000\,km/s. Of particular interest for space weather is that this halo originated on then back side of the Sun and the spectral observations from UCoMP clearly shows the receding nature of this halo CME, providing evidence that this CME is not propagating towards Earth. We emphasize the potential value of UCoMP \ion{He}{1} 1083 for space weather application and how, even minimally processed \ion{He}{1} 1083 images, can help differentiating between CMEs directed towards and away from the Sun.

For the 2021 July 31 prominence eruption shown in Figure~\ref{fig:amp_figure_20210730}, a twisted structure is present above the solar limb, which erupts shortly after the end of the UCoMP observations.  In this case, the the data does not show a significant asymmetry in the wavelength dependence of the intensity. Different parts of the  twisted prominence exhibit emission in differing wavelength bands, hinting at the complex multidirectional motion. This is a good example of the spectroscopic power of UCoMP  observations for disentangling the 3-dimensional twisting motions in solar eruptions, associated with the loss of stability of the flux rope system.

\begin{figure*}
    \centering
    \includegraphics[width=1.0\linewidth]{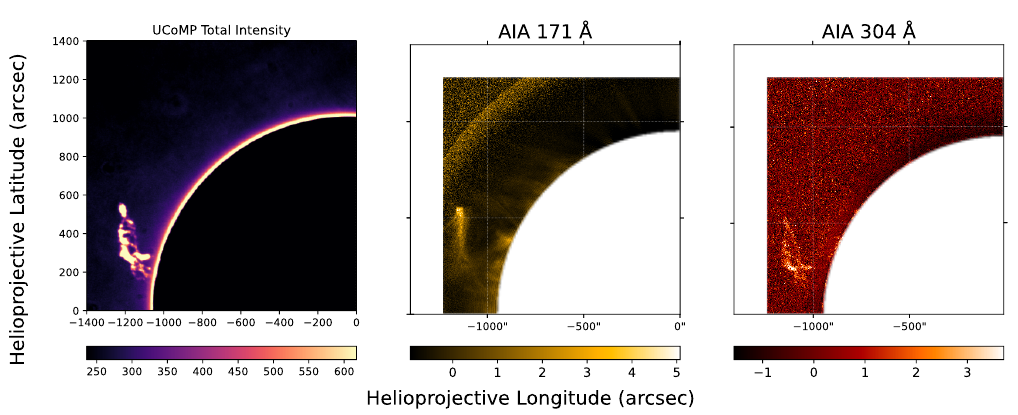}
    \caption{The left plot displays the maximal UCoMP spectral line intensity during a solar eruption on 2021 July 30 at 20:00 UT. The two plots on the right depict the same eruption captured by the AIA 171\,{\AA} and 304\,{\AA} filters with Normalizing Radial Gradient Filter applied.}
    \label{fig:UCoMP_AIA_Plot}
\end{figure*}

An important finding is that the bright prominence material is seen extending in the corona with brightness of greater than 1,000\,$\mu$B$_{\odot}$. The prominence material is a factor of 50 and up to 100 times brighter than the forbidden coronal lines regularly observed in the IR, making the \ion{He}{1} 1083 an easily detectable diagnostic of the erupting prominence material above the sky background. This was found to be the case for all major eruptions detected in the UCoMP data discussed throughout this work. Most notably, we did not detect a significant drop in the \ion{He}{1} 1083 intensity of the prominence material with its propagating outward away from the Sun. In the case of the 2021 July 15 eruption, the prominence material was seen clearly leaving the edge of the FOV, making the case for clear observability of erupting prominences in the \ion{He}{1} 1083 line in the middle corona. The  UCoMP \ion{He}{1} 1083 data provide a complementary spectral diagnostic with a larger FOV extent than AIA. This is especially notable for the example of the eruption on 2021 July 15, as shown in Figure~\ref{fig:K-Cor_AIA_UCoMP_Plot}, where the UCoMP observations clearly showed a prominence eruption associated with a halo CME that exhibited redshifts, providing us with a ready diagnostic of the receding CME.

\begin{figure*}[htbp]
    \centering
    \includegraphics[width=\linewidth]{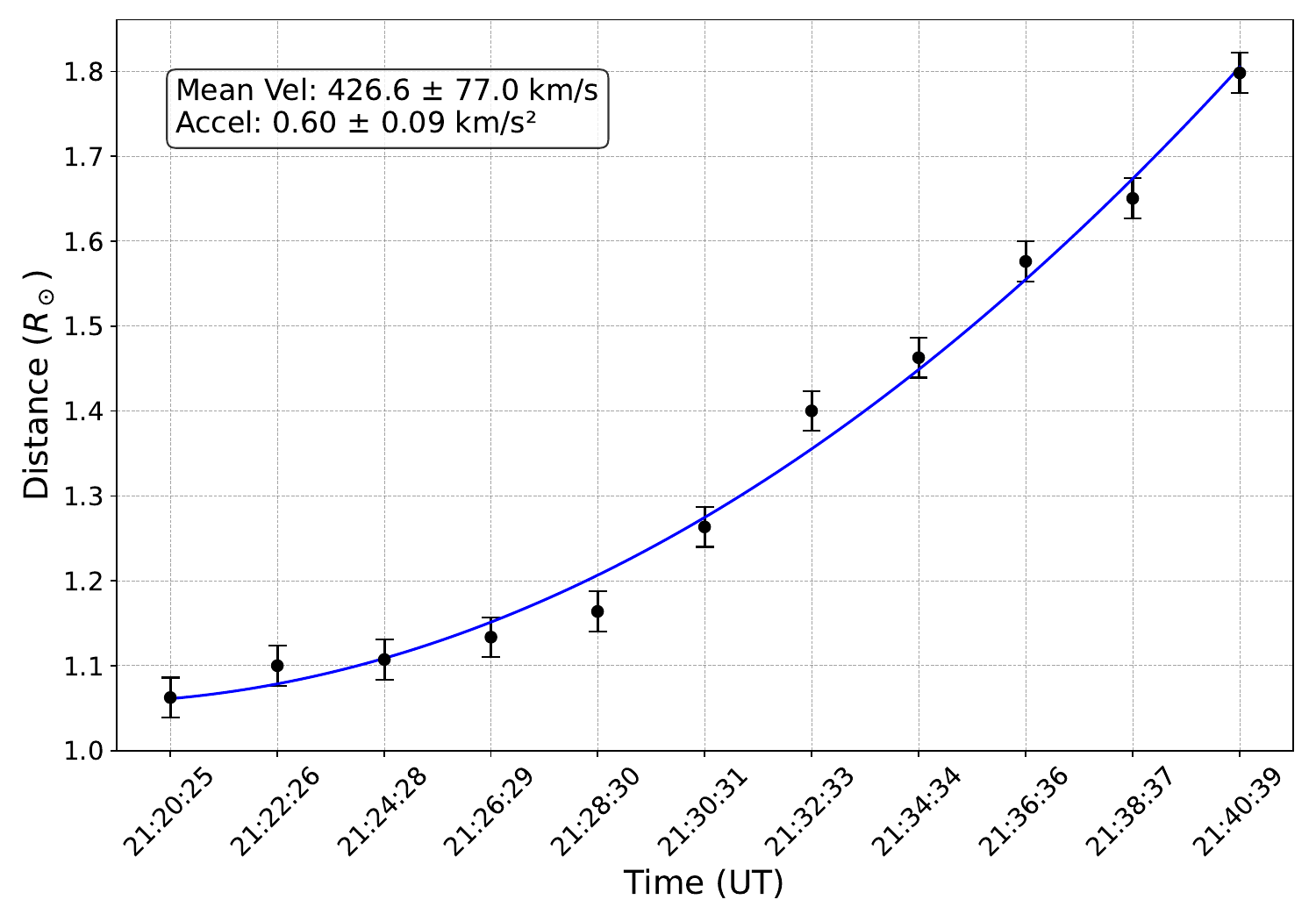}
    \caption{Evolution of the K-Cor white light data CME core distance above the limb on 2021 July 15. The error bars representing the uncertainty of the location of the CME front is derived from the uncertainty of its location from our algorithm, equating to about 4 pixels in the K-Cor images. The blue line represents a parabola fit, from which the mean velocity and acceleration during the interval of study, along with their uncertainties, are derived. }
    \label{fig:K-Cor_vel_plot}
\end{figure*}

\section{Discussion}
\label{sec:discussion}

The UCoMP \ion{He}{1} 1083 commissioning phase data provided us with a few clear examples of the promise of using the \ion{He}{1} 1083 as an erupting prominence diagnostic. The high brightness, observability up to 1.5\,R$_{\odot}$ and the potential for Doppler velocity measurements indicates the possibility of using this diagnostic for space weather applications if the saturation problem in bright eruptive prominences can be resolved and UCoMP data can be made available in near real-time. 
Future observations would require significantly more dense sampling of the spectral line, which is achievable with the instrument. This more complete sampling over a wider range of Doppler velocities would allow for a more robust Doppler velocity inference but would also result in a significantly longer time to complete a full line scan, which is not ideal for fast CMEs. For space weather applications, a balance between line sampling and data cadence must be worked out. 

Additionally, to establish if the \ion{He}{1} 1083 is an applicable diagnostic for space weather forecasting, we need to estimate what percentage of CMEs have associated \ion{He}{1} 1083 counterparts. We analyzed 79 days of UCoMP observations and identified solar eruptions on 14 of those days given that most days very few frames were taken sporadically. For comparison, the SOHO LASCO CME Catalog (\cite{cdaweventlist}) reports 31 eruptions during the same observing windows while the UCoMP was taking \ion{He}{1} data, indicating prominence eruptions associated with CMEs on roughly 45\% of events. A reason for this lesser than unity rate is the non-constant presence of cool material in CMEs~\citep{2025arXiv251209234G}, and most importantly, the significantly lesser than unity duty cycle of ground-based observatories. Thus, exact one-to-one agreement between the two instruments is not expected. Previous studies have also found a robust association of erupting prominences with CMEs~\citep{2003ApJ...586..562G,2025JApA...46...64D}.
These results imply that if UCoMP (and in the future large Coronagraph instrument in the COSMO suite~\citep{2023BAAS...55c.392T}) were observing continuously in the \ion{He}{1} 1083\ line, it could yield a significant CME detection rate. The relatively high correspondence nevertheless demonstrates that \ion{He}{1} 1083 is a robust diagnostic of CME activity and that UCoMP can reliably detect a significant fraction of eruptions identified by space-borne coronagraphs. Importantly, UCoMP provides a unique set of  ground-based observations, highlighting its potential as a complementary tool to other CME observing telescopes.

\section{Conclusions}
The primary objective of this study was to evaluate the observability of erupting prominences in the \ion{He}{1} 1083 data and the ability of the ground-based UCoMP telescope to detect them. Our findings indicate that erupting prominences are clearly (and quite brightly) seen with UCoMP in the \ion{He}{1} 1083, making this instrument a reliable tool for studying the evolution of CMEs. The \ion{He}{1} 1083 nm spectroscopic data can yield key plasma parameters, including line amplitude and an estimation of the Doppler velocities of the erupting prominences. This is a significant result as it provides a reliable, ground-based method for measuring the plasma parameters in erupting prominences entrailed in CMEs, with potential application  for space weather forecasting.
When combined with data from complementary instruments, such as SDO/ AIA and K-Cor, the UCoMP observations allow for the construction of a more detailed, three-dimensional picture of CME evolution through the low solar corona, as UCoMP allows for the radial velocity measurement of the prominence plasma detected in the \ion{He}{1} 1083 data. For some of these events, UCoMP provided a complimentary information of the early phases of the eruption, compared to the SDO/AIA observations.

\begin{acknowledgments}
The authors would like to thank Drs. Alfred de Wijn, Roberto Casini, and Stanislav Gunar for the valuable discussions that significantly improved this manuscript. We would like to express our gratitude to Dr. Joan Burkepile for proofreading the manuscript and providing the CME front propagation speed estimates. We thank Dr. Steven Tomczyk (Solar Scientific LLC) for the theoretical transmission profiles of the Lyot Filter of the UCoMP instrument. The NSF National Center for Atmospheric Research is a major
facility sponsored by the NSF under Cooperative Agreement No. 1852977. This work was supported
by NSF Grant Award 2504074 (SHINE). The K-Cor~\cite{KCor_reference} and UCoMP~\citep{UCoMP_reference} data are a courtesy of the Mauna
Loa Solar Observatory, operated by the High Altitude Observatory, as part of
the National Center for Atmospheric Research (NCAR). The following Python
packages were used in this work: \texttt{astropy}~\citep{astropy_2022};
\texttt{numpy}~\citep{Hunter:2007}; \texttt{scipy}~\citep{2020SciPy-NMeth}; and \texttt{sunpy}~\citep{sunpy_community2020}. This research has made use of the Astrophysics Data
System, funded by NASA under Cooperative Agreement 80NSSC21M00561.

\end{acknowledgments}

%

\vspace{5mm}




\bibliography{CHEESE_Eclipse_2024_results}{}
\bibliographystyle{aasjournal}



\end{document}